\documentclass{article}
\begin{document}
\begin{flushright}
FNT/T 99/19
\end{flushright}
\vskip 1cm
\begin{center}
{\LARGE LARGE-ANGLE BHABHA SCATTERING \\
\vskip 0.2cm
AND LUMINOSITY AT DA$\Phi$NE \footnote{Talk given by C.M. Carloni Calame at the
workshop PHYSICS AND DETECTORS FOR DA$\Phi$NE (Frascati, 16-19 November 1999), to
be published in the workshop proceedings.}}
\vskip 1cm
Carlo Michel Carloni Calame~$^{1,2}$, Cecilia Lunardini~$^{3}$, 
Guido Montagna~$^{1,2}$, Oreste Nicrosini~$^{2,1}$, Fulvio Piccinini~$^{2,1}$ \\
\vskip 0.5cm
{\em $^{1}$ Dipartimento di Fisica Nucleare e Teorica, Universita' di Pavia} \\
{\em $^{2}$ INFN, Sezione di Pavia} \\
{\em $^{3}$ SISSA and INFN, Sezione di Trieste} \\
\end{center}
\vskip 0.5cm
\baselineskip=11.6pt
\begin{abstract}
The accurate knowledge of luminosity at $e^+ e^-$ flavour factories 
requires the precision 
calculation of the Bhabha cross section at large scattering angles.
In order to achieve a theoretical accuracy at the 0.1\% 
level, the relevant effect of QED radiative corrections is taken into account in
the framework 
of the Parton Shower method, which allows exclusive event generation. On 
this scheme, a Monte Carlo event generator (BABAYAGA) is developed for data 
analysis. To test the reliability of the approach, a benchmark calculation, 
including exact $O(\alpha)$ corrections and higher-order leading logarithmic 
contributions, is developed as well, implemented in a Monte Carlo 
integrator (LABSPV) and compared in detail with the BABAYAGA predictions. The 
effect of initial-state and final-state radiation, $O(\alpha)$ next-to-leading 
and higher-order leading corrections is investigated and discussed in the 
presence of realistic event selections. The theoretical precision of BABAYAGA
is estimated to be at the $0.5\%$ level.
\end{abstract}
\baselineskip=14pt
\section{Introduction}
The precise determination of the machine luminosity 
is necessary for the successful accomplishment of the physics program of 
$e^+ e^-$ colliders operating in the region of the low-lying 
hadronic resonances, such as DA$\Phi$NE (Frascati)~\cite{daph}, VEPP-2M 
(Novosibirsk)~\cite{novos}, 
as well as for the BABAR~\cite{bfact,babar} and BELLE~\cite{bfact,belle}
experiments at PEP-II and KEKB. 
In particular, the precise measurement of the hadronic cross section requires a 
luminosity determination with a total relative error better than
1\%~\cite{achim}.
It is well known that the luminosity of $e^+ e^-$ colliders can be precisely 
derived by the relation $L = N / \sigma_{th}$, where $N$ and $\sigma_{th}$ are 
the number of events and the theoretical cross section of a given reference
reaction. In order to make the total luminosity error as small as
possible, the reference process should be characterized by a large cross section
and calculable with high theoretical
accuracy. At low-energy $e^+ e^-$ machines, the best candidate fulfilling the
above criteria is the Bhabha process ($e^+ e^-  \to e^+ e^-$) detected at large 
scattering angles.

On the theoretical side, precision calculations of the large-angle Bhabha (LABH)
cross-section are therefore demanded, with a theoretical accuracy at the 
${O}(10^{-3})$ level. This requires the inclusion in the calculation  
of all the relevant radiative 
corrections, in particular the large effects due to 
photonic radiation. The complete and exclusive simulation of events 
in generators is also strongly required by the experimental analysis.
\section{Theoretical approach}
The calculation of the 
Bhabha scattering cross-section, corrected by the effects of photon radiation, 
and the 
corresponding event generation is performed according to the 
master formula~\cite{prd}
\begin{eqnarray}
&&\sigma_{corrected}=\int dx_- dx_+ dy_- dy_+ \int d\Omega_{lab}
D(x_-,Q^2)D(x_+,Q^2) \times \nonumber\\ 
&&D(y_-,Q^2)D(y_+,Q^2) \frac{d\sigma_0}{d\Omega_{cm}}\big(x_-x_+s,
\theta_{cm}\big)
J\big(x_-,x_+,\theta_{lab}\big)\Theta(cuts) \, .
\label{eq:sezfs}
\end{eqnarray}   
In the previous equation, the electron Structure Function (SF) $D(x,Q^2)$ is the
solution of DGLAP equation in QED. It takes into account soft-photon 
exponentiation and multiple hard bremsstrahlung emission in the leading log
(LL) approximation~\cite{sfcoll}, both for the QED initial-state (ISR) and 
final-state radiation (FSR). The QED-DGLAP equation can be exactly solved by 
means of the QED Parton Shower (PS) algorithm~\cite{psqcd}, which allows also 
exclusive photon generation in the LL approximation. A more detailed discussion 
about the implementation of the PS algorithm as adopted in the present analysis
will be given elsewhere~\cite{noi}. In eq.~\ref{eq:sezfs}, 
$d\sigma_0 / d\Omega$ is the ``hard-scattering'' differential 
cross section relevant for centre of mass (c.m.) energy around 1 GeV, 
including the photonic $s$- and $t$-channel
diagrams, their interference and the (small) contribution due to 
$\Phi$ exchange. 
In the hard-scattering 
cross section, the correction due to vacuum polarization is taken into account
as well, according to the parameterization and the recipe 
given in ref.~\cite{vpol}. The effect of the 
running coupling constant at $\sqrt{s} \simeq M_\Phi$ is to enhance 
the cross section by 
$\approx 2(2.5)\%$ for $20^\circ(50^\circ) \leq 
\vartheta_{\pm} \leq 160^\circ(130^\circ)$. 
The jacobian factor $J\big(x_-,x_+,\theta_{lab}\big)$
in eq.~\ref{eq:sezfs} accounts for the boost from the c.m. to the 
laboratory frame due to emission by initial state $e^{+}$ and $e^{-}$ of 
unbalanced radiation, while $\Theta(cuts)$ stands 
for (arbitrary) experimental cuts implementation. 

Upon the above-sketched theoretical background, a new Monte Carlo
(MC) generator (BABAYAGA) for simulation of the LABH
process at $\Phi$-factories has been 
developed. In the program both ISR and FSR are simulated and the complete
kinematics of the generated events is reconstructed in the LL approximation. 
The possibility of performing an up to ${O}(\alpha)$ calculation 
of eq.~\ref{eq:sezfs} is included as well, in view of a comparison with the 
exact ${O}(\alpha)$ perturbative results.

In order to test the precision and the reliability of the Bhabha generator, an
exact ${O}(\alpha)$ calculation has also been addressed, by computing the 
up to $O(\alpha)$ corrected cross-section, consisting of 
soft+virtual~\cite{vs} and hard photon corrections~\cite{hard}.
Moreover, higher-order LL terms can 
be summed on top of the exact $O(\alpha)$ cross section whitin the collinear
SF approach, following the 
algorithm of ref.~\cite{a2l}. This formulation is 
available in the form of a MC integrator (LABSPV), which is a suitable 
modification of the SABSPV code described in ref.~\cite{sabspv}.
%
\begin{figure}[t]
 \vspace{9.0cm}
\includegraphics{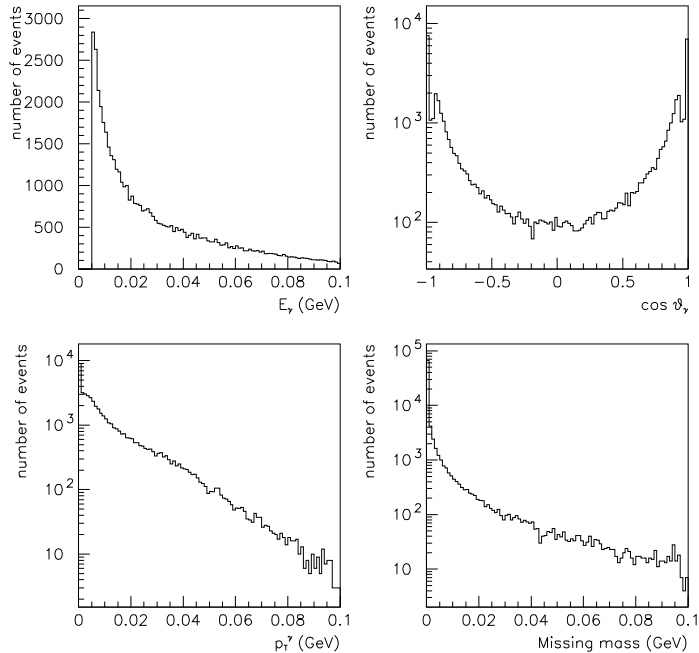}
\caption{\it Differential distributions obtained by means of BABAYAGA. 
Clockwise: energy, cosine of the angle, $p_{\perp}$ of the most energetic 
photon and missing mass.
\label{fig.1}}
\end{figure}
\section{Numerical results}
In the following, the selection criteria adopted for the analysis and the 
simulations correspond to realistic data taking at DA$\Phi$NE and 
VEPP-2M, at c.m. energy $\sqrt{s} = 1.019$~GeV. The energy cut imposed on the 
final-state electron and positron is $E_{min}^{\pm} = 0.4$~GeV, 
with the angular acceptance of $20^\circ \leq \vartheta_{\pm} 
\leq 160^\circ$ or $50^\circ \leq \vartheta_{\pm} 
\leq 130^\circ$ and the (maximum) 
acollinearity cut allowed to vary in the range
$\xi_{max} = 5^\circ$-$25^\circ$. 

A sample of simulations obtained by means of 
BABAYAGA is shown in fig.~\ref{fig.1}, where 
the energy, the cosine of the angle, the $p_{\perp}$ of the most
energetic photon of each event and the missing mass of the event are plotted. As
expected,
the behaviour of photonic radiation (soft and collinear to charged particles)
is well reproduced by the PS. The effect of FSR has also been investigated. 
The corrected cross section including only
ISR has been compared with the corrected cross section including 
both ISR and FSR. We noticed 
that, as a consequence of the rather severe
cuts, the total effects of photon radiation is to reduce the integrated cross
section by an $O(10\%)$ amount. As expected, half of the whole effect must 
be ascribed to FSR when non-calorimetric (``bare'')
event selection~\cite{a2l,cv} is adopted.
\begin{figure}[t]
 \vspace{8.0cm}
\includegraphics{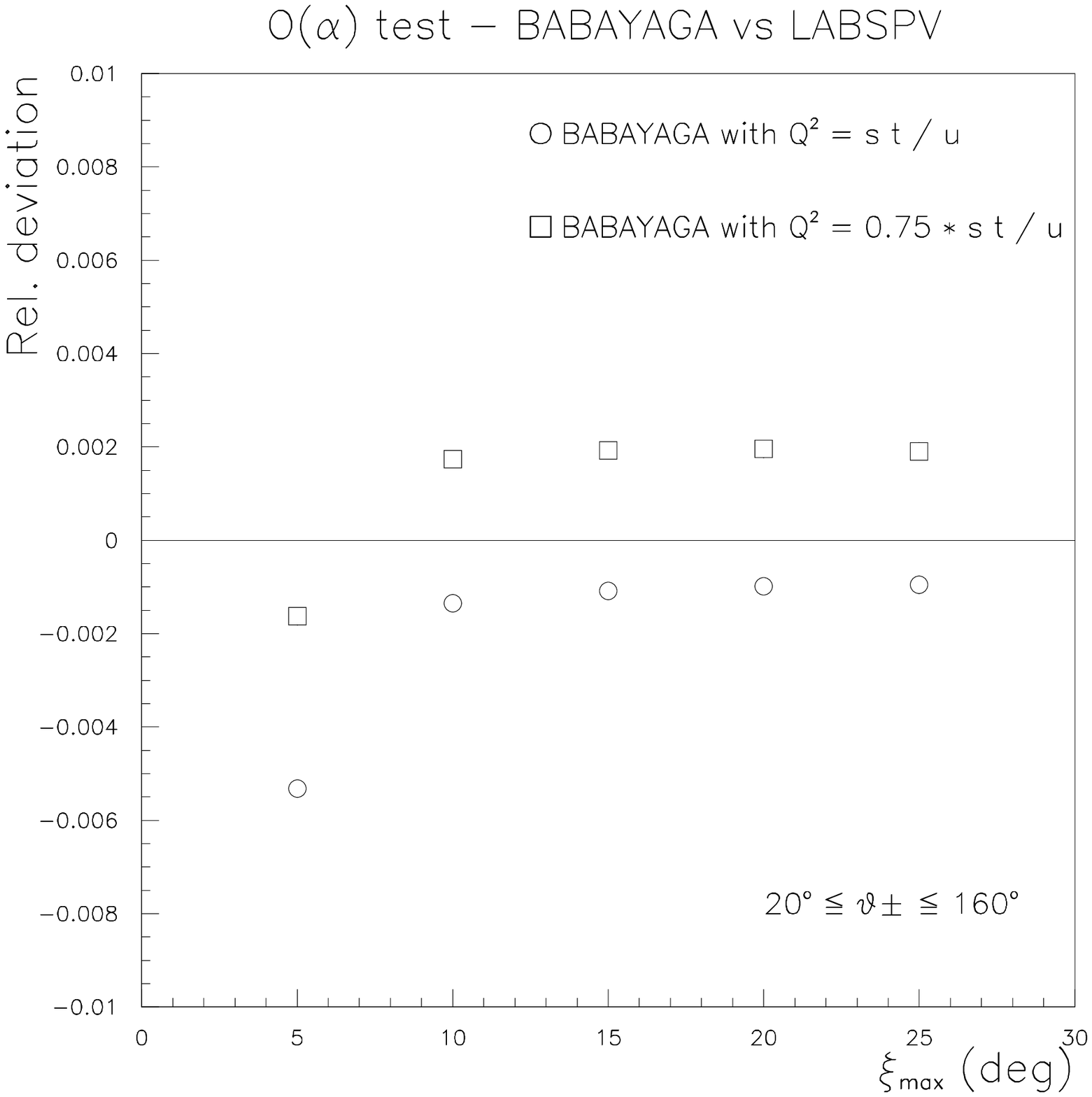}
\includegraphics{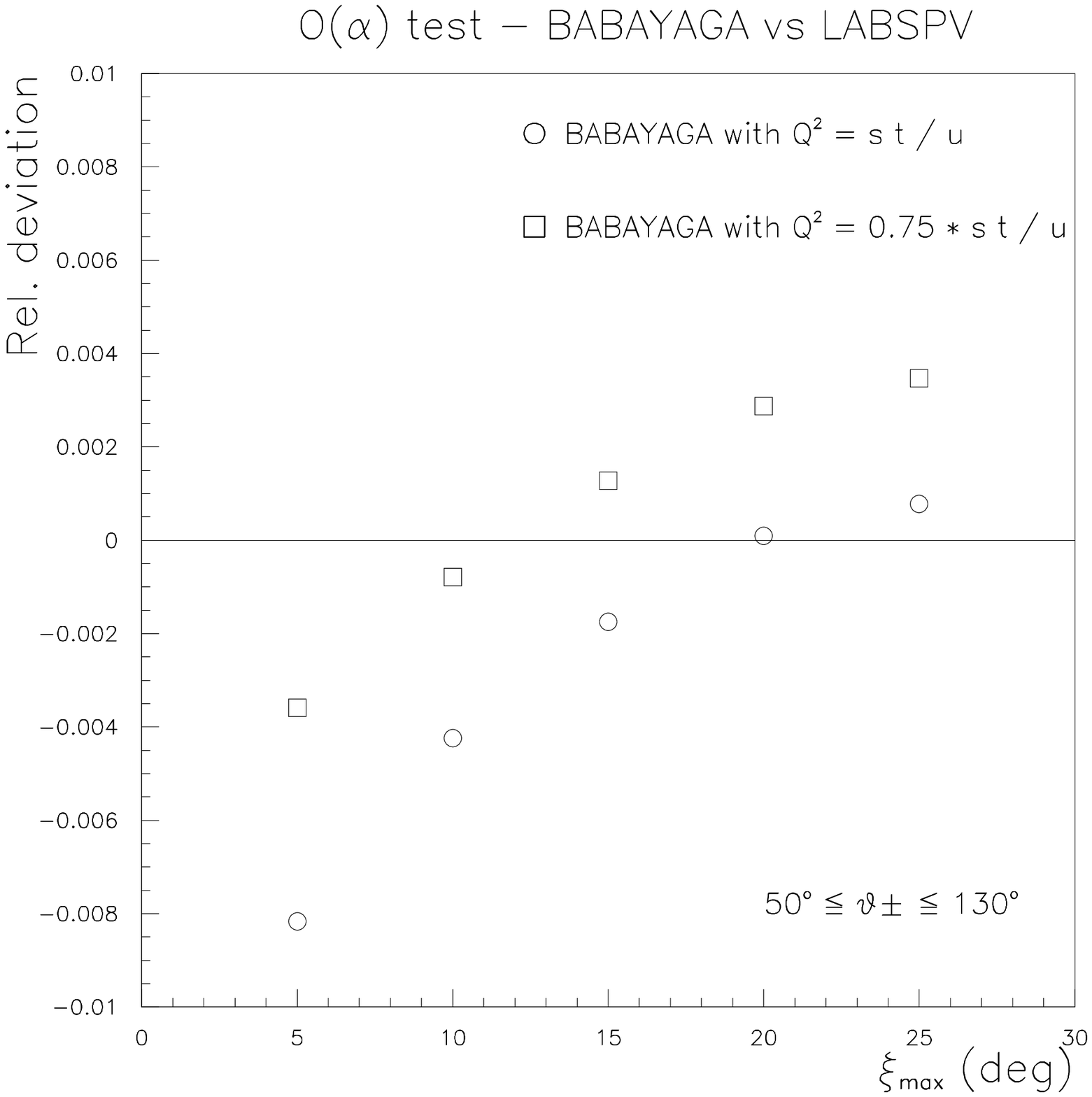}
 \caption{\it Relative differences 
between the exact ${O}(\alpha)$ and 
the up to ${O}(\alpha)$ PS cross sections, as functions 
of the acollinearity cut and for two choices of the scale 
in the electron SF's, in two different angular acceptance regions.
         \label{fig.2}}
\end{figure} \\
\indent
The comparison between the exact $O(\alpha)$ calculation and the $O(\alpha)$ 
predictions of the PS generator allows to evaluate the size of the $O(\alpha)$
next-to-leading-order (NLO) corrections missing in the LL approximation PS 
predictions. Moreover, this comparison can be a useful guideline 
to improve the agreement between perturbative and PS results, 
for example, by properly choosing the virtuality $Q^2$ 
in the electron SF in such a way that the bulk of $O(\alpha)$ NLO terms is 
effectively reabsorbed into the LL contributions. The scale choice $Q^2 = s t/u$
($s$, $t$ and $u$ are the usual Mandelstam variables) allows to keep under 
control the dominant structure due to initial-, final- and initial-final-state 
interference radiation~\cite{stu}. As a function of the 
acollinearity cut, the relative difference
between the exact $O(\alpha)$ cross section and the corresponding 
PS one is shown in fig.~\ref{fig.2}, for the angular acceptances
$20^\circ \leq \vartheta_{\pm}\leq 160^\circ$ and 
$50^\circ \leq \vartheta_{\pm}\leq 130^\circ$  and for 
two different choices of the $Q^2$ scale in the PS, {\it i.e.} 
$Q^2 = s t/u$ and $Q^2 = 0.75\cdot s t/u$.  
It can be seen that, with the scale 
$Q^2 = 0.75\cdot s t/u$, the difference between the exact ${O}(\alpha)$ 
calculation and the PS predictions is within 0.5\%. This naive example 
illustrates how, for a given selection criterion, the level of agreement   
can be substantially improved by a simple redefinition of 
the maximum virtuality of the 
electromagnetic shower. Going beyond this simple recipe would require a 
true merging between perturbative calculation and 
PS scheme, which is beyond the scope of the present analysis.
\begin{figure}[t]
 \vspace{9.0cm}
\includegraphics{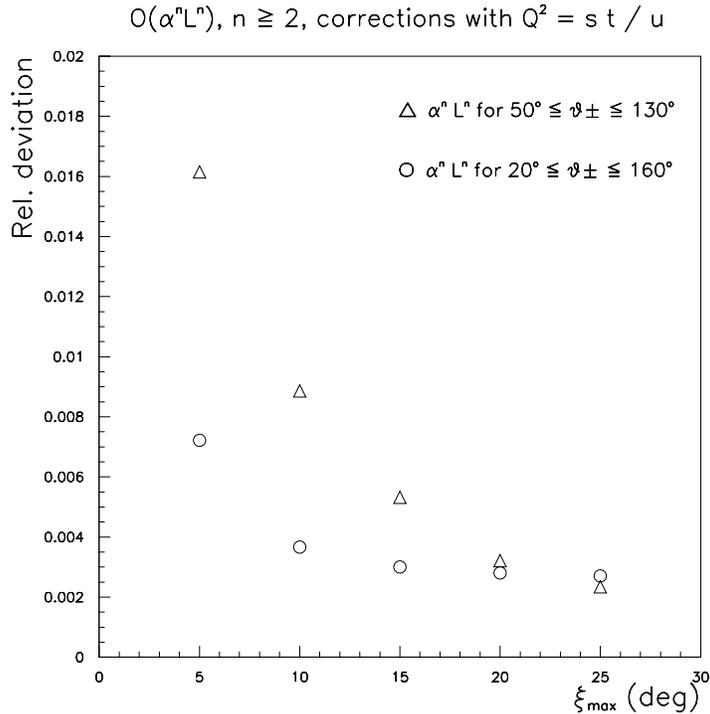}
 \caption{\it Relative deviation between $\cal O(\alpha)$ and all-orders
 corrected cross section, in the PS scheme, as a function of acollinearity cut.
 Two different angular acceptances for $e^{+}$ and $e^{-}$ are considered.
\label{fig.3}}
\end{figure}

In addition to the evaluation of the $O(\alpha)$ NLO 
corrections, it is important, for an assessment of the 
theoretical precision, to quantify 
the amount of the higher-order LL contributions with typical experimental cuts. 
The size of LL $O(\alpha^n L^n$) ($n \geq 2$) corrections can be derived
in the PS scheme by comparing the full all-order predictions 
with the corresponding up to $O(\alpha)$ truncation, as shown in 
fig.~\ref{fig.3}. The comparison shows that 
the $O(\alpha^n L^n$) corrections are unavoidable 
for a theoretical precision better than 1\%, being their contribution 0.7\% at 
$\xi_{max} = 5^\circ$ and 0.3-0.4\% for larger acollinearity cuts in the 
angular acceptance $20^\circ \leq \vartheta_{\pm} 
\leq 160^\circ$. The $O(\alpha^n L^n$) corrections are even more important, of 
the order of 1.5\%, in the narrower angular range $50^\circ \leq \vartheta_{\pm}
\leq 130^\circ$. It is worth noticing that the generators so far 
used by the experimental groups at Frascati and Novosibirsk
include only~\cite{cv} $O(\alpha)$ corrections, missing the important effect of
higher-order contributions.
\section{Conclusions and perspectives}
In order to provide predictions of interest for the luminosity determination at
$e^+e^-$ flavour factories, a precision calculation of the 
LABH process has been addressed. It is based on a QED PS algorithm (the details
of the formulation as adopted in the present paper will be given 
elsewhere~\cite{noi}), which accounts for corrections due to ISR and FSR (and
interference) in the LL approximation and allows the complete event generation.
A new MC event generator (BABAYAGA) has been developed and 
is available for a full experimental simulation; actually, it is under test at
Frascati and Novosibirsk. The overall
precision of the PS approach has been checked by means of a benchmark
calculation, which includes exact $O(\alpha)$ and higher-order LL corrections
and is available as a MC integrator (LABSPV), allowing for precise 
cross section calculations. Critical comparisons between the exact $\cal
O(\alpha)$ and the $\cal O(\alpha)$ PS calculations pointed out that the 
contribution of the $O(\alpha)$ NLO corrections is important for the required
theoretical precision. Moreover, the effect of higher-order $O(\alpha^n L^n)$ 
LL corrections has been evaluated to be at the 1-2\% level. 

By virtue of its generality, the PS approach could be employed to simulate and
to evaluate radiative corrections to other large-angle QED 
processes, as for example $e^{+}e^{-}\rightarrow \gamma\gamma$ or 
$e^{+}e^{-}\rightarrow \mu^{+}\mu^{-}$. 
An interesting application of PS would be the simulation of processes with 
tagged photons, e.g. $e^{+}e^{-}\rightarrow hadrons+\gamma$.

In conclusion, our analysis points out that
theoretical predictions aiming at a ${O}(10^{-3})$ precision must include the 
effects of both $O(\alpha)$ NLO terms and $O(\alpha^n L^n)$ LL 
contributions. As a consequence of that, we can estimate the present accuracy of
our generator BABAYAGA to be at 0.5\% level and the accuracy of the integrator
LABSPV at 0.1\% level. In the future, an improvement
of the presented approach is needed by means of an appropriate merging
of the exact $O(\alpha)$ matrix element with the exclusive photon
exponentiation realized by the PS algorithm. 
\section{Acknowledgements}
C.M. Carloni Calame is grateful to the organisers for the kind invitation 
and the pleasant atmosphere during the workshop. The authors
are indebted with A.~Bukin, G.~Capon, G.~Cabibbo, A.~Denig, S.~Eidelman, 
V.N.~Ivanchenko, F.~Jegerlehner, V.A.~Khoze, G.A. Ku\-kar\-tsev, 
J.~Lee-Franzini, G.~Pancheri, I.~Peruzzi, Z.A.~Silagadze and G.~Venanzoni for
useful discussions, remarks and interest in their work. 
The authors acknowledge partial support from the EEC-TMR Program, 
Contract N.~CT98-0169. C.M.~Carloni Calame and C.~Lunardini wish to thank the 
INFN, Sezione di Pavia, for the use of computer facilities.
\end{document}